\documentclass[10pt,a4paper]{report}

\usepackage{fancyheadings}
\usepackage{ifthen}
\usepackage{enumerate}
\usepackage{amssymb}
\usepackage[mathscr]{eucal}
\usepackage[errorshow]{tracefnt}
\usepackage{amsmath}
\usepackage{amssymb}
\usepackage{array}
\usepackage{amsthm}
\usepackage{eucal}
\usepackage{eufrak}
\usepackage{epsf}
\usepackage{epsfig}
\usepackage{pgf,pgfarrows,pgfautomata,pgfheaps,pgfnodes,pgfshade}
\usepackage[apple mac]{inputenc}
\usepackage{picins}

\def\onalign#1{\leavevmode\vtop{\baselineskip=0pt\lineskip=-1.2ex
\ialign{##\crcr#1\crcr}}}

\RequirePackage{hyperref}

\numberwithin{equation}{section}

\begin{document}

 \pagestyle{empty}

\begin{center}

\vspace*{2cm}

\noindent {\LARGE\textsf{\textbf{Introductory Lectures on Multiple Membranes}}}
\vskip 2truecm


\begin{center}
{\large \textsf{\textbf{Neil Copland}}} \\
\vskip 1truecm
     {\it   {Theoretische Natuurkunde, \\
     Vrije Universiteit Brussel \& International Solvay Institutes, \\
     VUB-campus Pleinlaan 2, B-1050, Brussel, Belgium,\\\vskip 1truecm
        e-mail:} {\tt ncopland@vub.ac.be}} \\
\end{center}
\vskip .5 cm

\small{Based on lectures given by the author at the Sixth
International Modave Summer School on Mathematical Physics, held in
Modave, Belgium, August 2010.}

\vskip 1 cm
\centerline{\sffamily\bfseries Abstract}
\end{center}

\noindent These lecture notes introduce the multiple membrane theories known as BLG and ABJM. We assume the reader is familiar with string theory, but not with M-theory, 11-dimensional supergravity or membranes. We therefore start with a background on M-theory and its extended objects before discussing BLG and ABJM. The link to string theory via dimensional reduction will be maintained throughout.

\newpage

\pagestyle{plain}
\tableofcontents
\chapter{Introduction}

These lectures aim to give an introductory overview of M-theory focussing on its fundamental objects: membranes and fivebranes, before going on to a more detailed look at the BLG and ABJM theories, which are believed to describe multiple membranes. This is clearly a vast subject and huge areas will not be discussed, for example in the discussion of M-theory there will be no discussion of matrix theory, and though I will always try to make clear the relation to string theory there will be no discussion of M-theory compactifications to four dimensions. Similarly the discussion of BLG (Bagger-Lambert-Gustavsson) and ABJM (Aharony, Bergman, Jafferis and Maldacena) will only have time to cover the basics and not huge amount of work than has been generated subsequently.

The first lecture will discuss the basics of M-theory, while the second will focus on branes in M-theory, especially membranes. Hopefully this will provide some context for the last two lectures which will cover BLG and ABJM respectively. There should not be too many specific prerequisites, other than a knowledge of string theory. There will, however, be some superspace expressions, though these will not be discussed in detail.

This is meant to be an introduction not a review, and though I have tried to site the major contributions to cite every related paper exhaustively is beyond the scope of these notes. Useful sources for more details and references  are, for the first lecture, Townsend\cite{Townsend:1996xj} for dualities and for some explicit calculations\cite{Miemiec:2005ry}. For branes in M-theory see the review by Berman\cite{Berman:2007bv}. For BLG and ABJM the best source is the original papers \cite{Bagger:2006sk,Bagger:2007jr,Bagger:2007vi,Gustavsson:2007vu,Aharony:2008ug} and also the review \cite{Klebanov:2009sg}.

\chapter{Introduction to M-Theory}

It is now 15 years since the existence of M-theory was discovered. I say the ``existence of M-theory was discovered," rather than ``M-theory was discovered'', because we still know so little about what M-theory actually is. M-theory was really born when it was realised that the strong coupling limit of type-IIA string theory is eleven dimensional, and that this eleven-dimensional theory's low energy limit was eleven-dimensional supergravity. Witten named this limit M-theory, and it came to be seen that all five string theories could be obtained from the one eleven-dimensional theory which is what is more commonly referred to as M-theory.

Of course the first question usually asked about M-theory is what does `M' stand for? It wasn't really specified at the time, and Witten himself says it stands for ``Magic, Mystery or Matrix,'' according to taste. Since then there have been many other suggestions put forward from ``Mother,'' to ``Murky, Muddled or Messy.'' Another popular suggestion is that the `M' is actually an upside down `W' for Witten.

A more sensible idea is that the `M' stands for membrane, because M-theory is a theory of membranes - which seem to play the role of strings - along with fivebranes, which are more akin to D-branes. A lot of what is know about M-theory is based on 11-dimensional supergravity and the membrane and fivebrane solutions of it. As we will see in the second lecture, we also have Lagrangians for a single membrane or fivebrane, but when we want to analyse stacks of co-incident M-theory branes our lack of knowledge of the fundamental theory comes into focus.

Another question that immediately occurs is why 11-dimensions? What's so good about it and why stop there? As usual a powerful tool that can lead to simple arguments is supersymmetry. If we're looking for a supersymmetric theory with no massless particles of spin greater than two then the maximum dimension allowed is 11. In fact, even before the advent of M-theory people were studying 11-dimensional supergravity as a possible origin of a unified theory; it is the maximal supergravity theory and all other supergravities can be derived from it by dimensional reduction. Eleven is also the maximal dimension to admit supersymmetric extended objects.

We have already mentioned that the theory contains membranes and fivebranes, and the supermembrane action was also known before the concept of M-theory existed. It can be doubly dimensionally reduced to the fundamental string in type IIA string theory. Of course it can also be directly dimensionally reduced to a D2 brane when we do not identify the compactified spacetime direction with one of the worldvolume directions. The fivebrane can also be reduced in two ways, and before the `D-brane revolution' this was seen as a drawback. Together with the pp-wave and Kaluza-Klein monopole solutions, which are also present in 11-dimensions, the membrane and fivebrane can be reduced to yield all the branes of IIA string theory. Using different compactifications there is a web of dualities relating the five different string theories amongst themselves and to this mysterious 11-dimensional M-theory.

So the five perturbative string theories are subsumed by unique non-perturbative M-theory. Obviously this is not true in a practical sense, and since we know so little about M-theory there is no need to give up on string theory quite yet! One is then lead to ask why there has been relatively little progress in understanding M-theory while so much is done on strings? Ultimately it's because it is hard! 

The most promising development in recent years has been the BLG (Bagger-Lambert-Gustavsson)\cite{Bagger:2006sk,Bagger:2007jr,Bagger:2007vi,Gustavsson:2007vu} and ABJM (Aharony-Bergman-Jafferis-Maldacena)\cite{Aharony:2008ug} theories. These are descriptions of field theories on coincident membranes. Knowing the actions and algebras of these membrane worldvolume fields should hopefully lead to insights into what the fundamental degrees of freedom are and answer questions about their counting. 

Ultimately we're still looking for answers to the biggest questions in M-theory, and that's one of the things that makes it so fascinating.

\section{Eleven-Dimensional Supergravity}

Much of the time when people say they are doing M-theory they are doing 11-dimensional supergravity, which takes a relatively simple form compared to lower dimensional supergravities (and the lower dimensional supergravities all follow from dimensional reduction). In fact obtaining other (especially 4-dimensional)supergravities was the main motivation for writing it down in the first place. This was done by Cremmer, Julia and Scherk\cite{Cremmer:1978km} in `78. The field content consists of the metric $g_{\mu\nu}$, a rank 3 anti-symmetric tensor field $C_{\mu\nu\rho}$ and a 32 component Majorana gravitino $\Psi^\alpha_\mu$. These have 44, 84 and 128 physical degrees of freedom respectively. (That is if we assume the $C$ field is transverse, i.e. we have invariance under the gauge transform $C\rightarrow C+d\Lambda$ where Lambda is a 2 -form). The Lagrangian is given by
\begin{eqnarray}
I_{11}&=&\frac{1}{16\pi G^{(11)}_N}\int d^{11}x\sqrt{-g^{(11)}}\left[R_{(11)}-\frac{1}{2.4!}G^2-\frac{1}{2}\bar{\Psi}_\mu\Gamma^{\mu\nu\rho}D_\nu(\Omega)\Psi_\rho\right.\nonumber\\ 
& &\left.-\frac{1}{192}\left(\bar{\Psi}_\mu\Gamma^{\mu\nu\rho\lambda\sigma\tau}\Psi_\tau+12\bar{\Psi}^\nu\Gamma^{\rho\lambda}\Psi^\sigma\right)G_{\nu\rho\lambda\sigma}\right]\nonumber\\ 
& & -\frac{1}{96\pi G^{(11)}_N}\int C\wedge G\wedge G +\mbox{terms quartic in} \ \Psi 
\end{eqnarray}
where $G=dC$ is the field strength of $C$ and $\Omega_\mu^{ab}$ is the spin connection, which appears in the covariant derivative $D_\nu(\Omega)\Psi_\rho=\left(\partial_\nu-\frac{1}{4}\Omega_\nu^{ab}\Gamma_{ab}\right)\Psi_\rho$. 

Note the Chern-Simons term for $C_{\mu\nu\rho}$, this allows membranes to couple to fivebranes. The equation of motion for the 3-form potential $C$ can be re-written in the form
\begin{equation}
d(*G+\frac{1}{2}C\wedge G)=0 \qquad 
\end{equation}
where $*G$ is the Hodge dual of $G$. This has the form of a Bianchi identity and we can identify $*G+C\wedge G/2$ with $dC^{(6)}$ where $C^{(6)}$ is a 6-form potential and the dual of $C$. The field strength of $C^{(6)}$ is $G^{(7)}=*G=dC^{(6)}-C\wedge G/2$. The appearance of $C$ in this field strength makes  reformulation of the action in terms of only the dual field strength difficult. The existence of 3- and 6-form potentials was suggestive of extended objects with 3 and 6 space-time dimensional worldvolumes, even before the ``D-brane revolution" in string theory.

Of course, as written this is just a quantum theory of gravity with all the standard problems of renormalisability. It is only the low-energy limit of M-theory, just as type IIA supergravity is the low energy limit of type II string theory.

\section{Ten Dimensions from Eleven: Type IIA Sring Theory}\label{sec:Mred}

We can compactify 11-dimensional supergravity on a circle of fixed radius in the $x^{10}=z$ direction\cite{Huq:1983im,Giani:1984wc}. The gorey details of this and other basic M-theory calculations can be found in \cite{Miemiec:2005ry}. From the 11-dimensional metric we obtain the 10-dimensional metric, a vector field and a scalar (the dilaton). The 3-form potential leads to both a 3-form and a 2-form in 10 dimensions. Using the Sherk-Schwarz reduction procedure the ansatz is
\begin{eqnarray} \label{nbc:dimred}
g^{11}_{ab}&=&e^{-2\phi/3}g_{ab}+e^{4\phi/3}C^{(1)}_aC^{(1)}_b\qquad\qquad\, C_{abc}=C^{(3)}_{abc}\nonumber\\
g^{11}_{az}&=&e^{4\phi/3}C^{(1)}_a \qquad\qquad\qquad \qquad\qquad\quad C_{abz}=B_{ab}\nonumber\\
g^{11}_{zz}&=&e^{4\phi/3}.
\end{eqnarray}
$g_{ab}$ is the 10 dimensional metric with $a,b,c,\ldots$ representing 10-dimensional indices. $C^{(1)}$, $B$ and $C^{(3)}$ are one, two and three forms respectively. $\phi$ is the dilaton and we have performed a Weyl rescalings on both the 11- and 10-dimensional metrics so that the resulting action is in the string frame. To perform the reduction first note that using the above ansatz with $\phi=0$ would allow us to use the original Kaluza-Klein reduction
\begin{equation}
R_{(11)}=R_{(10)}-\frac{1}{4}\left(G^{(2)}_{ab}\right)^2,
\end{equation}
where $G^{2}=dC^{1}$ and $R_{(10)}$, $R_{(11)}$ are the eleven-, ten-dimensional Ricci tensors respectively. To be able to use this we should before and after perform eleven- then ten-dimensional Weyl rescaling using
\begin{equation}
\tilde{g}_{\mu\nu}=e^{2\sigma}g_{\mu\nu}\implies \tilde{R}_{(d)}=e^{-2\sigma}\left[R_{(d)}-2(d-1)\Delta \sigma-(d-1)(d-2)\partial_\mu\sigma\partial^{\mu}\sigma\right]\, .
\end{equation}

The resulting bosonic action is 
\begin{eqnarray}
I_{10}&=&\frac{2\pi \onalign{\hidewidth$ \ell$\hidewidth\cr\cr$\mathchar'26$}_{pl}}{16\pi G^{(11)}_N}\int d ^{10}x\sqrt{-g}\left\{e^{-2\phi}\left[R_{(10)}+4(\partial \phi)^2-\frac{1}{2.3!}H_{abc}^2\right]\right. \nonumber\\&&\left.-\left[\frac{1}{4}\left(G^{(2)}\right)^2+\frac{1}{2.4!}\left(G^{(4)}\right)^2\right]\right\}-\frac{2\pi\onalign{\hidewidth$ \ell$\hidewidth\cr\cr$\mathchar'26$}_{pl}}{16\pi G^{(11)}_N}\int B\wedge G^{(4)}\wedge G^{(4)}.
\end{eqnarray}
We have compactified on a circle of radius $\onalign{\hidewidth$ \ell$\hidewidth\cr\cr$\mathchar'26$}_{pl}=\ell_{pl}/{2\pi}$ and $H$ and $G^{(4)}$ are the field strengths of  $B$ and $C^{(3)}$ respectively.

However, since the 11-dimensional metric is asymptotically flat we would also like the 10-dimensional metric to have this property. As things stand we have $g_{ab}\rightarrow e^{2\phi_o/3}\eta_{ab}$ as we go towards spatial infinity, where $\phi_0$ is the asymptotic value of the dilaton. We rescale the metric to an asymptotically flat form, and rescale other fields to remove extra factors of $e^{\phi_o}$. This requires
\begin{eqnarray}
g_{ab}\rightarrow e^{2\phi_0/3}g_{ab}\ \qquad\qquad&& C^{(1)}_a\rightarrow e^{\phi_0/3}C^{(1)}_a\nonumber\\
B_{ab}\rightarrow e^{2\phi_0/3}B_{ab}\qquad\qquad&& C^{(3)}_{abc}\rightarrow e^{\phi_0}C^{(3)}_{abc}.
\end{eqnarray}
Since $g_s$ (the IIA string coupling which counts loops in string amplitudes) is given by $g_s=e^{\phi_0}$, this leaves the action in the form
\begin{eqnarray}
I_{10}&=&\frac{g_s^2}{16\pi G^{(10)}_N}\int d ^{10}x\sqrt{-g}\left\{e^{-2\phi}\left[R(g)-4(\partial \phi)^2+\frac{1}{2.3!}H_{abc}^2\right]\right.\nonumber\\&& \left.-\left[\frac{1}{4}\left(G^{(2)}\right)^2+\frac{1}{2.4!}\left(G^{(4)}\right)^2\right]\right\}\nonumber\\ &&-\frac{g_s^2}{2.16\pi G^{(10)}_N}\int B\wedge G^{(4)}\wedge G^{(4)},
\end{eqnarray}
if we make the identification
\begin{equation}\label{nbc:consts}
G^{(10)}_N=\frac{G^{(11)}_N}{2\pi\onalign{\hidewidth$ \ell$\hidewidth\cr\cr$\mathchar'26$}_{pl} g_s^{2/3}}.
\end{equation}
This is precisely the type-IIA string theory low-energy effective action. We have fixed $z$ on a circle of radius $\onalign{\hidewidth$ \ell$\hidewidth\cr\cr$\mathchar'26$}_{pl}$, but the radius of the eleventh dimension measured at infinity is naturally measured in the 11-dimensional metric:
\begin{equation}\label{nbc:R11pre}
R_{11}=\frac{1}{2\pi}\lim_{r\rightarrow\infty}\int\sqrt{|g_{zz}|}dz=\onalign{\hidewidth$ \ell$\hidewidth\cr\cr$\mathchar'26$}_{pl} e^{2\phi_o/3}=\onalign{\hidewidth$ \ell$\hidewidth\cr\cr$\mathchar'26$}_{pl} g_s^{2/3}.
\end{equation}
This relation is extremely important in M-theory and it reduces (\ref{nbc:consts}) to the standard Kaluza-Klein form
\begin{equation}
G^{(10)}_N=\frac{G^{(11)}_N}{V_{11}},
\end{equation}
where $V_{11}=2\pi R_{11}$ is the volume of the internal space.
Standard formulae in 10 and 11 dimensions give us that $G_N^{(10)}=8\pi^6g_s^2(\alpha')^4$ and $G^{(11)}_N=\frac{( \ell_{pl})^9}{32\pi^2}$, so that (\ref{nbc:consts}) leads to the relation $\onalign{\hidewidth$ \ell$\hidewidth\cr\cr$\mathchar'26$}_{pl}= \ell_s g_s^{1/3}$. Thus we can write the following relations between the constants in 11-dimensions and those of IIA string theory:
\begin{eqnarray}
\ell_{pl}=2\pi \ell_s g_s^{1/3},\\
R_{11}=\ell_s g_s.\label{nbc:Mconsts}
\end{eqnarray}
We can see from this second relation that as we go to strong coupling we are going to the decompactification limit; i.e. towards the 11-dimensional theory. It is also useful to express $R_{11}$ in units of the 11-dimensional Planck length (divided by $2\pi$), allowing (\ref{nbc:R11pre}) to be rewritten as
\begin{equation}\label{nbc:R11}
R_{11}=g_s^{2/3}.
\end{equation}

\section{Ten Dimensions form Eleven: Other String Theories}

We will see how the F1-string and D2-, D4- and NS5-branes of type IIA string theory can be obtained from the 11-dimensional M2- and M5-branes. However, from a fundamental theory which unifies the five consistent string theories we should expect to find the complete complement of IIA branes, and the connection to the other four string theories should be clear. To complete the IIA picture the D0 and D6 branes are easily found from compactification. The D0-particle corresponds to one unit of the quantised momentum in the periodic 11th dimension with higher momentum states corresponding to coincident D0-particles (see also the following section). The D6-brane corresponds to the 11-dimensional Kaluza-Klein monopole\cite{Townsend:1995kk}.

Next we can relate to type IIB strings - we have become very used to the idea that T-duality relates the two type II string theories. Recall this duality relates IIA string theory compactified on a circle of radius R with IIB on a circle of radius $1/R$ under exchange of winding and momentum modes. It follows that IIB on a circle is equivalent to M-theory on $T^2$ under such an exchange. Letting $R_{11}$ and $R_{10}$ go to zero with a fixed ratio leads to uncompactified type IIB string theory with IIB string coupling $g_S^{(B)}=R_{11}/R_{10}$. The $SL(2,\mathbb{Z})$ symmetry of type IIB - which includes the S-duality that relates weak to strong coupling ($g_S^{(B)}\leftrightarrow 1/g_S^{(B)}$) - is just the $SL(2,\mathbb{Z})$ of reparameterisations of the torus\cite{Townsend:1996xj}. Note that the chiral type-IIB theory comes from the non-chiral 11-dimensional theory, something that had previously been forbidden by `no-go' theorems. The chirality is introduced by massive spin-2 multiplets coming from the membrane ``wrapping" modes on $T^2$\cite{Bergshoeff:1995as}.

The heterotic string seems is a more difficult proposition to obtain, given that it has different numbers of left-movers and right-movers on the worldsheet. However, by compactifying a five-brane on the 2-complex-dimensional surface $K3$ (which has topology such that it admits 19 self-dual and 3 anti-self-dual 2-forms) one gets $(19,3)$ scalars from the 2-form, $(0,8)$ Fermions and $(5,5)$ other scalars, exactly what one would expect on the 10-dimensional heterotic string worldsheet\cite{Townsend:1995de}. One can also get the $E_8\times E_8$ heterotic string in 10 dimensions by compactifying M-theory on $\mathbb{R}^{10}\times S^1/\mathbb{Z}_2$\cite{Horava:1995qa}. Here again we obtain a chiral theory from a non-chiral one. This time previous `no-go' theorems are circumvented by compactifying on an orbifold rather that a manifold. Since $E_8\times E_8$ heterotic and $SO(32)$ heterotic are T-dual to one another, once we have the connection to one we can quickly find connections to the other.

Type I string theory comes from orbifolding type IIB and through similar arguments to those above it can be deduced that type I string theory (or rather its T-dual, type IA) is the  $R\rightarrow 0$ limit of M-theory on a cylinder of radius $R$\cite{Horava:1995qa}. We are seeing that we are dealing with a moduli space of vacua that is in general 11-dimensional, with 10-dimensional perturbative string expansions only in certain 10-dimensional limits.

Dualities in M-theory lead to dualities in various  lower dimensions. Membrane-five-brane duality in 11 dimensions leads to string-string duality in 6 dimensions, between fundamental F-strings and solitonic D-strings, both in the heterotic theory\cite{Duff:1996rs}. However, as in the previous section, compactifying different string theories on different manifolds can lead to a duality between the heterotic string and the IIA string\cite{Hull:1994ys}. Further compactification of each theory on $T^2$ leads to a more surprising duality of dualities: the solitonic string has a non-perturbative S-duality which is a perturbative T-duality in the dual fundamental string picture\cite{Duff:1994zt}. Compactification leads to amazing symmetries revealing themselves in the lower dimensional theories. Compactification to extremely low dimensions leads to the appearance of exceptional Lie and affine algebras. An interesting question is whether these hidden symmetries like $E_{10}$ or $E_{11}$ are in some way fundamental in 11-dimensions.

\section{Eleven Dimensions from Ten}

Previously we obtained the link between M-theory and type IIA by looking at the low energy theory and the membrane action upon dimensional reduction on a circle. However the real spark that led to the study of M-theory was the realisation that when you took the strong coupling limit of type-IIA string theory you were naturally led to an 11-dimensional theory.

The key to the argument is the presence of D0-branes in type-IIA. They are stable excitations with mass given by $1/(\ell_s g_s)$ in the string frame. As the mass diverges as $g_s\rightarrow 0$ they are non-perturbitive, so we are testing beyond perturbation theory. On the other hand, at strong coupling their mass goes to zero. They also carry a conserved $U(1)$ charge. The only way these states could be interpreted in the strong coupling limit was as the first level of the Kaluza-Klein tower coming from the compactification of 11-dimensional supergravity on a circle (with higher Kaluza-Klein modes corresponding to bound states of D0-branes).

If we start from the 11-dimensional supergraviton multiplet which is massless  ($M^2_{11}=-p_\mu p^\mu=0$) then after compactification on a circle $M^2_{10}=-p_a p^a=p_{11}p^{11}$ and the mass is given by the momentum in the eleventh dimension. Since this is periodic, the momentum is quantised as $p_{11}=n/R_{11}$ for integer $n$, and the mass of the first excitations is $1/R_{11}$, which on comparison with the D0-brane mass gives us back (\ref{nbc:Mconsts}). Once again we see that strong coupling is the decompactification ($R_{11}\rightarrow\infty$) limit.

When these inferences were made it was before the `D-brane revolution' and D0-branes were not really understood, however it was known there was a Ramond-Ramond gauge field and an associated central charge in the supersymmetry algebra. Further, there were the related charged BPS states. Being BPS they were in short multiplets and protected at strong coupling where there became infinitely man of them. The Kaluza-Klein interpretation was the only one possible.

We could also look at the supersymmetry of the low energy theory. As we go from weak to strong coupling in type IIA  we maintain IIA supersymmetry so we cannot go to any of the other string theories, and the coupling to the short centrally charged multiplets leads us to the dimensional reduction of 11-dimensional supergravity as above.

For IIB there is no central charge and the massless states remain the same (massive multiplets necessarily contain states of spin 4, and massless spin 4 states are not consistent) as we go to strong coupling. The conclusion is that the low energy theory at strong coupling is also type-IIB supergravity and we have the S-duality mentioned in the previous section.

\chapter{Branes in M-Theory}

In this lecture we will discuss the extended objets of M-theory, the membrane and the fivebrane. The membrane is fundamental, and is thought to play a role like the fundamental string in string theory: it reduces to the string after dimensional reduction, and open membranes can end on fivebranes much like strings on a D-brane. Unfortunately the analogy cannot be taken much further, no-one knows how to quantise the membrane (indeed any p-brane with $p>1$) and it has a continuous spectrum. Further, we don't have a picture of what the degrees of freedom are in M-theory as we do in string theory. There we are used to the picture of light strings stretching between branes becoming massive as the branes coincide. Labeling the strings by which branes they end on leads to the familiar $U(N)$ gauge theory with $N^2$ degrees of freedom for $N$ branes. We will see later that for $N$ coincident membranes there are $N^{3/2}$ degrees of freedom, and $N^3$ for coincident fivebranes. Interpreting this remains a great challenge.

There are two main perspectives we can take on extended objects such as the membrane and fivebrane. We can look at them as solutions of 11-dimensional supergravity (these solutions will also have near horizon limits) and look at the field theories on their worldvolumes. This is at the heart of the AdS/CFT correspondence. The degrees of freedom on the worldvolume are goldstone modes from broken symmetries, including supersymmetries. Requiring that the Bosonic and Fermionic degrees of freedom match to give a supersymmetric worldvolume theory puts very strong constrains on the allowed extended objects, importantly the maximal dimension this can occur in is 11. Here the 8 scalars from broken translations in the directions transverse to the brane match with 8 Fermions from the broken supersymmetry. A fivebrane thus has only 5 scalars but will still have 8 Fermions if it preserves half the supersymmetry. The three additional Bosonic degrees of freedom come from broken gauge symmetries of the three-form $C$. This leads to a 2-form with anti-self-dual field strength on the fivebrane worldvolume. This makes the fivebrane worldvolume theory difficult to formulate. While the single membrane and fivebrane cases (which dimensionally reduce to Abelian gauge theory) have been known for some time, the multiple brane cases (which would have non-Abelian dimensional reductions) have proved elusive. For multiple membranes there has been much excitement (and papers) generated by the BLG and ABJM theories, which we will discuss in the last two lectures. We will discuss membrane and fivebrane supergravity solutions, single brane worldvolume actions and dimensional reductions.

\section{The M2-Brane as a Supergravity Solution}

Objects with three-dimensional worldvolumes were investigated as long ago as 1962 by Dirac\cite{Dirac:1962iy}. The 3-form potential of 11-dimensional supergravity is suggestive of coupling to such a membrane and an extremal membrane solution of 11-dimensional supergravity was found by Duff and Stelle\cite{Duff:1990xz}, taking the form
\begin{eqnarray}\label{nbc:M2sug}
ds^2&=&H^{-2/3}\eta_{\mu\nu}dx^\mu  dx^\nu +H^{1/3}\delta_{pq}dy^p dy^q, \nonumber\\
C&=&\pm \frac{1}{3!}H^{-1}\epsilon_{\mu\nu\rho}dx^\mu dx^\nu dx^\rho, \qquad \mbox{where}\ H=1+\left(\frac{R}{\rho}\right)^6.
\end{eqnarray}
The indices are split into $\mu,\nu,\ldots=0,1,2$ and $p,q,\ldots=3,4,\ldots,10$ and $\rho=\sqrt{\delta_{pq}y^py^q}$ is the transverse radius. $H$ has the harmonic property $\delta^{pq}\partial_p\partial_qH=0$. Everything but the form of $H$ follows from the killing spinor equation, the form is fixed by the equations of motion. While this solution preserves half the supersymmetry and saturates a BPS bound, it is not a soliton of the theory; the equations of motion are singular on the membrane and require a $\delta$-function source. This is the same behaviour as the string solution of supergravity found by Dabholkar, Gibbons, Harvey and Ruiz-Ruiz\cite{Dabholkar:1990yf}, which it reduces to under dimensional reduction.

\section{Worldvolume Action for a Single Membrane}

The supermembrane action in 11-dimensions was constructed by Bergshoeff, Sezgin and Townsend\cite{Bergshoeff:1987cm}. It is common to see just the Bosonic expression, as to write the fermionic parts in a compact manner utilises superspace. The differential $\Pi^A=dX^A-i\bar{\theta}\Gamma^m d\theta$ (where $\theta$ is the Fermionic co-ordinate and the spacetime index $A=0,\ldots,10$) is invariant under spacetime supersymmetries $\delta\theta=\epsilon$, $\delta X^a=-i\bar{\epsilon}\Gamma^{\mu}\theta$, and is used to construct the action. In Howe-Tucker form the action is then given by
\begin{equation}\label{nbc:M2wvol}
S=-\frac{1}{2}\int d^3\xi \left(\sqrt{-\gamma}\gamma^{ij}\Pi_i^{A}\Pi_j^{B}\eta_{AB}+\epsilon^{ijk}B_{ijk}-\sqrt{-\gamma}\right).
\end{equation}
$i$ labels worldvolume co-ordinates $0,1,2$ with metric $\gamma_{ij}$ of signature $(-,+,+)$and $B$ is 
\begin{equation}\label{nbc:M2b}
B_{ijk}=i\bar{\theta}\Gamma_{AB}\partial_i\theta\left[\Pi_i^{A}\Pi_j^{B}+i\Pi_i^A\left(\bar{\theta}\Gamma^B\partial_j\theta\right)-\frac{1}{3}\left(\bar{\theta}\Gamma^A\partial_i\theta\right)\left(\bar{\theta}\Gamma^B\partial_j\theta\right)\right]\, .
\end{equation}

Supersymmetry of the action follows from the `4$\psi$'s identity' which says a certain product of 4-arbitrary spinors vanishes in 4,5,7 or 11 dimensions (this is related to the 4 division algebras\cite{Baez:2010ye}). There is also a Fermionic kappa-symmetry that means half of the Fermionic degrees of freedom are redundant and can be gauge fixed. We will use the Bosonic part of the action later. 

\section{Superstring from a Membrane}

The membrane can be reduced to a superstring by double dimensional reduction, this is where the worldvolume dimension is reduced by one along with the spacetime dimension. This was performed by Duff, Howe, Inami and Stelle\cite{Duff:1987bx}. To make things tractable let's just deal with the Bosonic sector of the supermembrane action, which including the coupling to the the background 3-form $C$  is given by

\begin{equation}
S=\int d^3\xi \left(\frac{1}{2}\sqrt{-
\hat{\gamma}}\hat{\gamma}^{ij}\partial_i X^{\hat{m}}\partial_jX^{\hat{n}}\hat{g}_{\hat{m}\hat{n}}-\frac{1}{6}\epsilon^{ijk}\partial_iX^{\hat{m}}\partial_jX^{\hat{n}}\partial_kX^{\hat{p}}C_{\hat{m}\hat{n}\hat{p}}-\frac{1}{2}\sqrt{-\hat{\gamma}}\right),
\end{equation}
where $\hat{\gamma}^{ij}$ is the worldvolume metric and $g_{\hat{m}\hat{n}}$ is the background metric - 11-dimensional indices also wear hats.

Splitting the co-ordinates as $\xi^i=(\sigma^a,\rho)$ for $a=1,2$ and $X^{\hat{m}}=(X^m,z)$ for $m=0,1,\dots,9$ we make the gauge choice $z=\rho$ and demand $\partial_\rho X^m=0$, $\partial_z \hat{g}^{\hat{m}\hat{n}}=0$ and $\partial_z C_{\hat{m}\hat{n}\hat{p}}=0$. We can then make a reduction ansatz equivalent to (\ref{nbc:dimred})
\begin{eqnarray}
\hat{g}_{mn}&=&e^{-2\phi/3}g_{mn}+e^{4\phi/3}C^{(1)}_mC^{(1)}_n,\qquad\qquad C_{mnp}=C^{(3)}_{mnp}\, ,\nonumber\\
\hat{g}_{mz}&=&e^{4\phi/3}C^{(1)}_a, \qquad\qquad\qquad\qquad\qquad\quad\, C_{mnz}=B_{mn}\, ,\nonumber\\
\hat{g}_{zz}&=&e^{4\phi/3},
\end{eqnarray}
which implies that $\sqrt{-\hat{g}}=\sqrt{-g}$. It can be shown that substitution into the field equations leads to the string equation of motion one would expect from 
\begin{equation}
S=\int d^2\sigma \left(\frac{1}{2}\sqrt{-\gamma}\gamma^{ab}\partial_a X^{m}\partial_bX^{n}g_{mn}-\frac{1}{2}\epsilon^{ij}\partial_a X^{m}\partial_b X^{n}B_{mn}\right).
\end{equation}
($C^{(3)}, C^{(1)}$ and $\phi$ have decoupled here but persist in the Fermionic sector.) The $X^z$ component of the equations of motion yields an identity which confirms consistency. In fact substituting into the action directly gives a 2-dimensional action equivalent to that of the string. This can be extended to the full supersymmetric case which yields the superspace action of the type IIA superstring coupled to IIA supergravity. Since the IIA superstring is known to be a consistent quantum theory this gives hope that there should be a theory of membranes in 11-dimensions which is also consistent. Notice that the membrane is not conformally invariant but leads to the conformally invariant superstring, carefully following this through shows that the Weyl-transform of the string is a remnant of 11-dimensional diffeomorphism invariance.

\section{D2-Brane from a Membrane}

Rather than double dimensional reduction we can perform direct dimensional reduction by compactifying on a circle in one of the transverse directions to the membrane. This leads to a D2-brane in type- IIA string theory, in fact this was how the Fermionic part of the D-brane actions were first obtained\cite{Townsend:1995af}. Looking at the field content of the membrane and D2-brane theory we see that the only difference is that the on the D2-brane one of the scalars is replaced by a vector gauge field, these are dual to each other in 3-dimensions so we just need to implement this duality. We relabel the 8th worldvolume scalar $\varphi$ and promote $L=d\varphi$ to an independent worldvolume 1-form. We must then also impose $dL=0$ by the Lagrange multiplier term $AdL$. We can then eliminate L by its equation of motion and the action is now in terms of $F=dA$. The action (\ref{nbc:M2wvol}) becomes
\begin{eqnarray}
S&=&-\frac{1}{2}\int d^3\xi \left(\sqrt{-\gamma}\left[\gamma^{ij}\Pi_i^{m}\Pi_j^{n}\eta_{mn}+\frac{1}{2}\gamma^{ik}\gamma^{jl}\hat{F}_{ij}\hat{F}_{kl}-1\right)\right]\nonumber\\&&\qquad\qquad-\frac{1}{2}\int d^3\xi\epsilon^{ijk}\left[b_{ijk}+i\left(\bar{\theta}\Gamma_{11}\partial_i\theta\right)\hat{F}_{jk}\right]\, ,
\end{eqnarray}
where
\begin{equation}
\hat{F}_{ij}=F_{ij}-b_{ij}
\end{equation}
and
\begin{eqnarray}
\epsilon^{ijk}b_{ijk}&=&i\epsilon^{ijk}\bar{\theta}\Gamma_{mn}\partial_i\theta\left[\Pi_j^{m}\Pi_k^{n}+i\Pi_j^m\left(\bar{\theta}\Gamma^n\partial_k\theta\right)-\frac{1}{3}\left(\bar{\theta}\Gamma^m\partial_j\theta\right)\left(\bar{\theta}\Gamma^n\partial_k\theta\right)\right]\nonumber\\
&&+\left(\bar{\theta}\Gamma_{m}\Gamma_{11}\partial_i\theta\right)\left(\bar{\theta}\Gamma_{11}\partial_j\theta\right)\left(\partial_k X^m-\frac{i}{2}\bar{\theta}\Gamma^{m}\partial_k\theta\right)\, ,\\
\epsilon^{ijk}b_{ij}&=&-i\epsilon^{ijk}\bar{\theta}\Gamma_{m}\Gamma_{11}\partial_i\theta\left(2\partial_j X^{m}-i\bar{\theta}\Gamma^m\partial_j\theta\right)\, .
\end{eqnarray}
Note that $b_{ij}$ is the two-from appearing in the Wess-Zumino term in the Green-Schwarz superstring action. Its derivative $h=db$ is superinvariant, which allows $\hat{F}$ to be, so that we have a supersymmetric action. Kappa-symmetry can also be demonstrated. Couplings to background fields ($g_{ij}, B_{ij},\phi$) can also be introduced but things become even more complicated, things look more familiar if we just include Bosonic pieces giving
\begin{equation}
S=-\frac{1}{2}\int d^3\xi e^{-\phi}\sqrt{-\gamma}\left[\gamma^{ij}g_{ij}+\frac{1}{2}\gamma^{ik}\gamma^{jl}\left(F_{ij}-B_{ij}\right)\left(F_{kl}-B_{kl}\right)-1\right]\,.
\end{equation}

From this general form of the supersymmetric $p$-brane action was deduced. Alternatively, starting from the D2-brane and reversing the process illustrates the hidden 11-dimensional Lorentz invariance of string theory.

\section{The M5-Brane Solution to Supergravity}

The five-brane solution was first found in supergravity by Gueven\cite{Gueven:1992hh}, takes a similar form to that of the membrane
\begin{eqnarray}
ds^2&=&H^{-1/3}\eta_{\mu\nu}dx^\mu  dx^\nu +H^{2/3}\delta_{mn}dy^m dy^n,\nonumber\\
G&=&*_y dH,\qquad\mbox{where}\ H=1+\left(\frac{R}{\rho}\right)^3.
\end{eqnarray}
In defining $G$ we have used $*_y$, the Hodge star in the transverse directions. Again the indices are split, into $\mu,\nu,\ldots=0,1,\ldots, 5$ and $m,n,\dots=6,7,\ldots,10$ and $\rho=\sqrt{\delta_{mn}y^my^n}$ is the transverse radius.

The membrane is an ``electric" singular solution to the supergravity equations coupled to a membrane source. It has a Noether electric charge given by 
\begin{equation}
Q=\frac{1}{\sqrt{2}}\int_{S^7}(*G+\frac{1}{2}C\wedge G)=\sqrt{2}\kappa_{11}T_3.
\end{equation}
The five-brane, however, is a solitonic solution with  topological magnetic charge given by 
\begin{equation}
P=\frac{1}{\sqrt{2}\kappa_{11}}\int_{S^4}G=\sqrt{2}\kappa_{11}T_6.
\end{equation}
These charges obey a higher dimensional analogue of Dirac quantisation given by $QP=2\pi n$ for integer $n$, or equivalently $2\kappa_{11}^2 T_3T_6=2\pi n$. Along with the relation $T_6=\frac{1}{2\pi}T_3^2$, which can be deduced from the quantisation of the periods of $C$, this implies we have only one independent dimensionful parameter in 11 dimensions.

\section{The M5-Brane Worldvolume Action and Reduction}

There are difficulties formulating the worldvolume action for a five-brane as it contains a 2-form tensor field with anti-self-dual field strength. It is part of a $(0,2)$ tensor multiplet on the worldvolume, giving a superconformal theory in 6-dimensions. To write down an action approaches can be taken: one is to introduce an auxiliary field to ensure that the generalised self-duality condition appears as an equation of motion\cite{Pasti:1997gx,Bandos:1997ui}, and the other is to formulate the action in such a way that 6-dimensional general covariance is not manifest\cite{Perry:1996mk,Aganagic:1997zq}. Alternatively one can work without an action and use the equations of motion obtained via the superembedding formalism\cite{Howe:1997fb,Howe:1996yn}.

A starting point for deriving the action with non-manifest covariance was ensuring the correct dimensional reduction to a four-brane. This made the covariance in five of the dimensions obvious, but to prove it in the fifth spatial direction required more work. We single out the $x^5$ direction as different and write the indices $\hat{\mu}=(\mu,5)$. The anti-self-dual field is represented by $B_{\mu\nu}$ which is a 5-dimensional anti-symmetric tensor with 5-dimensional curl $H_{\mu\nu\rho}=3\partial_{[\mu}B_{\nu\rho]}$ and dual $\bar{H}^{\mu\nu}=\frac{1}{6}\epsilon^{\mu\nu\rho\lambda\sigma}H_{\rho\lambda\sigma}$. The metric also splits into $G_{\mu\nu}$, $G_{\mu5}$ and $G_{55}$, with $G_5$ being the 5-dimensional determinant. The Bosonic Lagrangian can then be written as
\begin{equation}
L=-\sqrt{-\mbox{det}(G_{\hat{\mu}\hat{\nu}}+iG_{\hat{\mu}\rho}G_{\hat{\nu}\lambda}\bar{H}^{\rho\lambda}/\sqrt{-G_5})}-\frac{1}{4}\bar{H}^{\mu\nu}\partial_5B_{\mu\nu}+\frac{1}{8}\epsilon_{\mu\nu\rho\lambda\sigma}\frac{G^{5\rho}}{G^{55}}\bar{H}^{\mu\nu}\bar{H}^{\lambda\rho},
\end{equation}
note the Born-Infeld and Wess-Zumino like terms. 

In the PST approach\cite{Pasti:1997gx,Bandos:1997ui} $B$ has additional $B_{\mu 5}$ components and there is an auxiliary field, $a$. However there are also extra gauge freedoms and one can set $B_{\mu5}=0$ and make a simple choice for $a$ so that the action becomes equivalent to the above. Both versions of the action can be supersymmetrised into a kappa-symmetric form. 

Similarly to the membrane case, double dimensional reduction on a circle gives a IIA string theory object, here a four-brane. At first a four-brane with an anti-symmetric tensor field is found, but analogously to the membrane-D2 reduction there is a worldvolume duality transformation that yields the standard D4-brane action with a worldvolume vector field\cite{Aganagic:1997zk}. Direct dimensional reduction leads to the NS5-brane.

%

\section{M-Brane Intersections and Open Membranes}

The membranes described previously do not have to be closed, they can have a boundary\cite{Strominger:1995ac}. The membrane couples to the 3-form $C$ whose field strength $G=dC$ is invariant under $C\rightarrow C+d\Lambda$ for some 2-form $\Lambda$. However, in the presence of a boundary the minimal coupling of $C$ to the membrane leads to a term $\int_{\partial M}\Lambda$. This would break gauge invariance, but if we couple the boundary (which will be a string) to a 2-form field which varies under gauge transformations as $b\rightarrow b-\Lambda$ we can preserve the gauge invariance. Of course the five-brane worldvolume contains exactly such a 2-form and we deduce that membranes can end on five-branes, making five-branes act much like the D-branes of M-theory. The five-brane worldvolume contains a string soliton\cite{Perry:1996mk,Howe:1997ue} to be identified with the end of a membrane. Parallels can be drawn to the D1-D3 intersection where the endpoint of the D1-string is a monopole and there is an intriguing duality between the different branes perspectives of the configuration, related to the ADHMN construction of monopoles. We will touch on the M-theory generalisation of this later.

\section{Coincident Brane Degrees of Freedom}

As stated before, for $N$ coincident membranes the number of degrees of freedom scales like $N^{3/2}$, while for fivebranes the scaling is $N^3$. There are three ways to obtain these relations\cite{Berman:2007bv}. The first is from brane thermodynamics\cite{Klebanov:1996un}: branes have horizons and we can apply the usual laws of black hole thermodynamics to get a temperature and entropy. This thermal entropy measures the degrees of freedom of the system and will depend on $N$, we demonstrate this for the membrane case.

Recall that in (\ref{nbc:M2sug}) we took
\begin{equation}
H=1+\left(\frac{R_{M2}}{\rho}\right)^6
\end{equation}
where we have included the subscript now to distinguish from the fivebrane. The harmonic function call alternatively be written
\begin{equation}
H=1+2^5\pi^2Q_{M2}\left(\frac{\ell_p}{\rho}\right)^6,
\end{equation}
which will give the result $Q_{M2}$ when you integrate the flux over the sphere at infinity, thus $Q_{M2}$ is  quantised to be an integer. The near horizon limit for the membrane is to take $\ell_p\rightarrow 0$ and $\rho\rightarrow 0 $ with $U=(2^5\pi^2Q)^{1/2}\rho^2/\ell_p^3$ fixed which gives $AdS_4\times S^7$ in terms of the new variable $U$. Again using (\ref{nbc:M2sug}) we see that
\begin{equation}
R_{AdS_4}=\frac{1}{2}(2^5\pi^2)^\frac{1}{6}N^\frac{1}{6}\ell_p\, ,
\end{equation}
which is twice the radius of the $S^7$ factor. 

From general properties of asymptotically AdS black holes, for large horizon, 
\begin{equation}
T\sim \frac{r_h}{R_{AdS_4}\ell_p},
\end{equation}
where $r_h$ is the horizon radius. We also have the standard Bekenstein-Hawking entropy
\begin{equation}
S=\frac{A}{4G_N}\, ,
\end{equation}
where $G_N$ is the {\it four}-dimensional Newton's constant, obtained by dividing the eleven dimensional one by the volume of the seven-dimensional sphere. Thus $G^{(4)}_N\sim G^{(11)}_N/(R_{S^7})^7$. Given that in 4 dimensions the area of the black hole horizon scales like $r_h^2$ rhe result is that the entropy scales like
\begin{equation}
S_{M2}\sim R_{M2}^9T^2\sim N^{3/2}T^2.
\end{equation}
The M5 case proceeds similarly and in fact the entropy has the same dependence on $R$, which leads to 
\begin{equation}
S_{M5}\sim R_{M5}^9T^5\sim N^{3}T^5.
\end{equation}

A second method one can use to obtain the $N^{3/2}$ scaling is low-energy scattering\cite{Witten:1998zw}. One looks at the low-energy fluctuations of a graviton in the background of the brane solution and calculates their absorption cross-section, which will again scale with the number of branes and give another measure of the degrees of freedom of the system. This gives the same scaling as above, which may not be surprising given the relation between the black hole entropy and horizon area, so you can debate whether this is an independent check\cite{Das:1996we}.

A third method is only available in the case of the fivebrane. This is because it involves anomalies and the membrane worldvolume, being odd-dimensional, is automatically anomaly free. Fivebrane anomaly cancellation for a single fivebrane is a very nice story involving inflow from a term added to the action and more subtle issues involving characteristic classes and the Chern-Simons term in the 11-dimensional supergravity Lagrangian. We again we recover the $N^3$ scaling, essentially because the the $C\wedge G\wedge G $ Chern-Simons term scales like $Q_{M5}^3$. The anomaly is a measure of degrees of freedom as it is in the same multiplet as the Weyl anomaly which gives the central charge. These anomaly arguments are reviewed in \cite{Berman:2007bv}.

\chapter{The BLG Theory}

In this lecture we will introduce the the Bagger-Lambert-Gustavsson theory which was proposed to describe multiple membranes. We start with some of the precursors to the theory before describing the work of Bagger and Lambert, and its equivalence to the work of Gustavsson. We will describe how this membrane theory can be related to multiple D2-brane theory via a novel Higgs mechanism, and some outstanding problems with the theory.

\section{Towards a Multiple Membrane Theory}

For a long time a theory of multiple membranes proved elusive. For coincident D-branes we have a intuitive picture of the degrees of freedom as massless strings stretching between two of the branes. This leads to $N^2$ degrees of freedom and we can work with matrix valued degrees of freedom. This is great -  we know what we're doing with matrices.

When it comes to membranes we have no picture of what the microscopic degrees of freedom are. Membranes are fundamental in M-theory, but how could open membranes give the $N^{3/2}$ degrees of freedom of coincident membranes or the $N^3$ for coincident fivebranes? And it seems we will no longer be able to work with familiar matrix algebras and commutators. Understanding membranes is difficult, or rather, interesting!

One of the main lines of attack is to proceed by analogy with string theory systems, since we know that the dimensional reduction of membranes should give better understood D-branes much can be learned. A useful system in this regard has been that of coincident D1-strings ending on D3-branes. This is of interest as there is a `duality' between the D1-string or D3-brane worldvolume pictures of this system. It is also mathematically interesting, especially with the appearance of the Nahm equation,
\begin{equation}
\frac{dX^{i}}{d\sigma} = \frac{i}{2}\epsilon_{ijk}[X^{j},X^{k}] \, .
\end{equation}
The Nahm equation is central to the ADMHN construction of monopoles. Here it appears as a BPS equation in the worldvolume theory of the D1-branes, while the endpoint of the D1-branes appears in the D3-brane theory as a monopole.

The M-theory analogue of this is coincident membranes terminating on a fivebrane. Basu and Harvey proposed that the BPS equation for this system should be\cite{Basu:2004ed}
\begin{equation}
\frac{dX^i}{d s}+\frac{M_{11}^3}{8\pi\sqrt{2N}}\frac{1}{4!}\epsilon_{ijkl}[G_5,X^j,X^k,X^l]=0.
\end{equation}
The main point is the bracket on the right, which is trilinear and antisymmetric. The $G_5$ should be taken as being part of the definition of the bracket. Once you have such an equation you can start thinking about what sort of action can produce it as a BPS equation, and what sort of fields $X^i$ should appear in it. In the original Basu and Harvey solutions these are co-ordinates on a fuzzy sphere representing the directions transverse to the membranes. 

\section{Associators and 3-Brackets}

A three-bracket structure seems to occur naturally in M-theory. While string theory has the Neveu-Schwarz 2-form, the membrane couples to the 3-form $C$. Since commutators are obtained from quantising Poisson-brackets, we can try to quantise higher dimensional analogues of this, namely the Nambu-Poisson bracket. Although this has been investigated in the mathematical literature it seems there is no canonical way to do this, with in particular odd brackets proving more complex the even ones. The bracket featured in the Basu-Harvey equation above is antisymmetrised across the 4 entries which are taken to be in some subalgebra of $U(N)$. $G_5$ is a particular matrix obeying  $G_5^2=1$. Of course $U(N)$ would not give the requisite degrees of freedom. However, there is a restriction to a certain subalgebra corresponding to representations on the fuzzy 3-sphere which does (fuzzy spheres being a discretisation of ordinary sphere's where the co-ordinates are represented by matrices). This subalgebra is not closed under multiplication, hence it is necessary to impose a projection after multiplication. This renders multiplication non-associative. This non-associativity is not without precedent in studies of the membranes, for example it is believed to appear on a membrane worldvolume in background C-field, much like non-commutativity appears on a D-brane in background B-field\cite{Hofman:2001zt}. Note that non-associativity is a breakdown of the Jacobi identity.

When Bagger and Lambert began constructing a theory for multiple membranes\cite{Bagger:2006sk} they took as a starting point a 3-bracket based on non-associativity. With a non-associative product $``\cdot"$ you can define the associator
\begin{equation}
<X^I,X^J,X^K> = (X^I\cdot X^J)\cdot X^K-X^I\cdot (X^J\cdot X^K)
\end{equation}
and an anti-symmetric 3-bracket by 
\begin{equation}
[X^I,X^J,X^K]= \frac{1}{2\cdot 3!}<X^{[I},X^J,X^{K]}>.
\end{equation}
While this was a starting point for the theory's construction, we will see that it is not necessary for its definition, we can define it in terms of conditions on the bracket only, or indeed the theory can be reformulated without reference to 3-brackets at all.

\section{The Bagger-Lambert Lagrangian}

We recall that along with the eight scalars on the worldvolume we have eight fermionic degrees of freedom. Thus supersymmetry leaves no room for any dynamical gauge field, unlike the D-brane case. Schwarz\cite{Schwarz:2004yj} considered the possibility that a Chern-Simons gauge field was the missing ingredient for the membrane theory, but was unable to find a consistent theory as he considered standard $U(N)$ valued fields. When trying to supersymmetrise the scalar-spinor sector of the 3-bracket theory the SUSY transforms close up to something which looks like a novel gauge transform which can be written in terms of a 3-bracket. Gauging this symmetry then supersymmetrising leads to a Lagrangian with a Chern-Simons gauge field\cite{Bagger:2007jr} as we shall see.

First we must give more details of the non-associative algebra, ${\cal A}$, with which we work (henceforth referred to as a 3-algebra, though more correctly a 3-Lie algebra, they have appeared before in the mathematical literature, originally in \cite{Filippov}). As well as the 3-bracket we need some kind of trace form, $\mbox{Tr}: {\cal A}\times{\cal A}\to {\mathbb C}$ which has the symmetry and invariance:
\begin{equation}
\mbox{Tr}(A,B)=\mbox{Tr}(B,A)\qquad \mbox{Tr}(A\cdot
B,C) = \mbox{Tr}(A,B\cdot C).
\end{equation}
The algebra should also be endowed with an analogue of complex conjugation, which we denote by $\#$, such that $\mbox{Tr}(A^\#,A)\ge 0$ for any $A\in {\cal A}$
(with equality if and only if $A=0$). 
These properties imply
\begin{equation}
\mbox{Tr}([A,B,C],D) =-\mbox{Tr}(A,[B,C,D]). \label{nbc:inv}
\end{equation}
In fact we can take this relation as fundamental without any reference to non-associativity if we wish.

For D-branes the supersymmetry algebra closes up to a gauge transformation (and a translation). Attempting to close the membrane scalar-spinor sector BL found closure up to a gauge like transformation which could be written in terms of a 3-bracket. The global version has the form
\begin{equation}
\delta X = [\alpha,\beta,X],
\end{equation}
with $\alpha,\beta \in {\cal A}$.
We require that this symmetry acts as a derivation on the 3-bracket, that is
\begin{equation}
\delta([X,Y,Z])= [\delta X,Y,Z]+[X,\delta Y,Z]+[X,Y,\delta Z].
\end{equation}
This implies
\begin{equation}
[\alpha,\beta,[X,Y,Z]]=[[\alpha,\beta,X],Y,Z]+[X,[\alpha,\beta,Y],Z]
+[X,Y,[\alpha,\beta,Z]].
\label{nbc:FI}
\end{equation}
This is known as the fundamental identity and for a Lie algebra where $\delta X = [\alpha,X]$ the equivalent would be the Jacobi identity. This fundamental identity appeared in earlier discussions of 3-brackets.

We expand ${\cal A}$ in terms of a basis $T^a$ which we assume to be Hermitian (i.e. $(T^a)^\#=T^a$).  Thus $X =X_aT^a$, $a = 1,..., n$, where $n$ is the dimension of $\cal A$ we can introduce structure constants
\begin{equation}
[T^a,T^b,T^c] = f^{abc}{}_{d}T^d,
\end{equation}
We can also use the trace-form to provide a metric
\begin{equation}
h^{ab} = \mbox{Tr}(T^a,T^b).
\end{equation}
For now we assume $h^{ab}$ is positive definite, and we can use it to raise indices like so: $f^{abcd}=h^{de}f^{abc}{}_{e}$.
The invariance of the trace (\ref{nbc:inv}) gives us that
\begin{equation}
f^{abcd} = -f^{dbca},
\end{equation}
and combined with the antisymmetry of the bracket we see
$f^{abcd} =f^{[abcd]}$. In terms of the structure constants the fundamental identity
(\ref{nbc:FI}) can be written
\begin{equation}
f^{efg}{}_{d}f^{abc}{}_{g}=f^{efa}{}_{g}f^{bcg}{}_{d}
+f^{efb}{}_{g}f^{cag}{}_{d}+f^{efc}{}_{g}f^{abg}{}_{d}.
\end{equation}

We assume there is an element $T^0$
that associates with everything such that
$f^{0ab}{}_{d}=0$. This decouples and
was interpreted as the centre-of-mass coordinate.

We can now consider the transformation
\begin{equation}
\delta X_d = f^{abc}{}_{d}\Lambda_{ab}X_c.
\end{equation}
In order to gauge this we introduce a covariant derivative
\begin{equation}
(D_\mu X)_a = \partial_\mu X_a - \tilde A_{\mu}{}^b{}_a X_b,
\end{equation}
where $\tilde A_{\mu}{}^{b}{}_a\equiv f^{cdb}{}_{a}A_{\mu cd}$. Although $\tilde A_{\mu}{}^{b}{}_a$ is not conventional gauge field the expressions look familiar. To ensure covariance of the derivative we need 
\begin{eqnarray} \delta \tilde A_{\mu}{}^b{}_a &=&\partial_\mu \tilde
\Lambda^b{}_{a} -\tilde \Lambda^b{}_{c}\tilde A_{\mu}{}^c{}_a +
\tilde A_{\mu}{}^b{}_c \tilde \Lambda^c{}_a\\
&= & D_\mu \tilde \Lambda^b{}_a.
\end{eqnarray}
The field strength
\begin{eqnarray}
\tilde F_{\mu\nu}{}^b{}_a  &=&\partial_\nu \tilde A_{\mu}{}^b{}_a  -
\partial_\mu \tilde A_{\nu}{}^b{}_a-\tilde A_{\mu}{}^b{}_c\tilde A_{\nu}{}^c{}_a
+ \tilde A_{\nu}{}^b{}_c \tilde A_{\mu}{}^c{}_a .
\end{eqnarray}
has the usual definition
\begin{equation}
\tilde F_{\mu\nu}{}^b{}_a X_b=([D_\mu,D_\nu]X)_a .
\end{equation}

With two 3-algebra indices, $\tilde A_{\mu}{}^{b}{}_a$ is in the space of linear maps from $\cal A$ to itself. Thus it acts as an element of $gl(n)$, in fact the anti-symmetry of $f^{abcd}$ restricts the symmetry algebra to within $so(n)$.

Bagger and Lambert were now able to construct a supersymmetry algebra which closed on the scalars up to one of these new gauge transformations. Closure on all the fields, including the gauge field fixes the supersymmetry algebra and the equations of motion. These could be shown to follow from the relatively simple Lagrangian
\begin{eqnarray}\label{nbc:BLLag}
\nonumber {\cal L} &=& -\frac{1}{2}(D_\mu X^{aI})(D^\mu X^{I}_{a})
+\frac{i}{2}\bar\Psi^a\Gamma^\mu D_\mu \Psi_a
+\frac{i}{4}\bar\Psi_b\Gamma_{IJ}X^I_cX^J_d\Psi_a f^{abcd}\\
&& - \frac{1}{2.3!}\mbox{Tr}([X^I,X^J,X^K],[X^I,X^J,X^K])\\
&&+\frac{1}{2}\varepsilon^{\mu\nu\lambda}(f^{abcd}A_{\mu
ab}\partial_\nu A_{\lambda cd} +\frac{2}{3}f^{cda}{}_gf^{efgb}
A_{\mu ab}A_{\nu cd}A_{\lambda ef}).
\end{eqnarray}
This Lagrangian contains no free parameters, the structure constants could be rescaled but due to the Chern-Simons term they are quantised. The Chern-Simons term involving the structure constants $f^{abcd}$ is non standard. The $\hat{A}_\mu{}^b{}_a=A_{\mu cd}f^{cdb}{}_a$ are considered the physical fields and appear in the gauge transformation and supersymmetry rules, though the Chern-Simons terms is written in terms of  $A_{\mu ab}$, it is invariant under shifts of $A_{\mu ab}$ that leave $\hat{A}_\mu{}^b{}_a$  invariant.

It is a worthwhile exercise to check that the Lagrangian is invariant under the above gauge transformations and the supersymmetry transformations
\begin{eqnarray}\label{nbc:BLsusy}
\nonumber \delta X^I_a &=& i\bar\epsilon\Gamma^I\Psi_a\\
\delta \Psi_a &=& D_\mu X^I_a\Gamma^\mu \Gamma^I\epsilon -\frac{1}{6}
X^I_bX^J_cX^K_d f^{bcd}{}_{a}\Gamma^{IJK}\epsilon \\
\nonumber \delta\tilde A_{\mu}{}^b{}_a &=& i\bar\epsilon
\Gamma_\mu\Gamma_IX^I_c\Psi_d f^{cdb}{}_{a}.
\end{eqnarray}

With a Lagrangian in place we can now ask about solutions. The simplest possible 3-algebra has four generators, and once normalised we must have
\begin{equation}
f^{abcd}\propto\epsilon^{abcd},
\end{equation}
which can be seen to satisfy the fundamental identity. This algebra is referred to as ${\cal A}_4$ and the gauge algebra is $so(4)$. It can be realised from a non-associative algebra based on the simplest fuzzy three-sphere algebra which is realised using the $SO(4)$ gamma matrices.

\section{Relation to Gustavsson's Formulation}

In parallel with Bagger and Lambert's work, Gustavsson also developed an algebra for multiple membranes and was eventually able to show closure of his supersymmetry algebra\cite{Gustavsson:2007vu}. His was based on a kind of graded algebra that had two different subspaces, $\cal A$ and $\cal B$. The gauge field lies in $\cal B$ while the scalars and fermions are in $\cal A$ The algebra requires three different bilinear brackets. Letting $\alpha,\beta \in {\cal A}$ and $A,B\in {\cal B}$ then we introduce
\begin{eqnarray}
\langle\alpha ,\beta \rangle &=& -\langle\beta,\alpha \rangle \in {\cal B}\, , \\
(A,\alpha) & \in& {\cal A} \, ,\\
\left[A,B\right]&=&-[B,A]\in {\cal B} \, ,
\end{eqnarray}
and require they obey the `associative condition'
\begin{equation}
(\langle\alpha,\beta\rangle,\gamma) =
(\langle\beta,\gamma\rangle,\alpha),
\end{equation}
as well as the the Jacobi-like identities 
\begin{eqnarray}
\langle (A,\alpha),\beta\rangle - \langle (A,\beta),\alpha\rangle  &=& [A,\langle \alpha,\beta\rangle ]\, ,\\
(A,(B,\alpha)) -(B,(A,\alpha))&=&([A,B],\alpha) \, ,\\
\left[\left[A,B\right],C\right]+\left[B,\left[A,C\right]\right] &=& [A,[B,C]]\, .
\end{eqnarray}
The last of these is the actual Jacobi identity and tells us that $\cal B$ is a Lie algebra. With this in place we can define an anti-symmetric triple product on $\cal A$ via
\begin{equation}
[\alpha,\beta,\gamma]:=(\langle\alpha ,\beta \rangle,\gamma).
\end{equation}

To show equivalence of the two constructions Bagger and Lambert\cite{Bagger:2007vi} first showed that the three bracket defined in this way by Gustavsson satisfies the fundamental identity (\ref{nbc:FI}). Conversely, if we start with an antisymmertic three-bracket satisfying the fundamental identity we can also reproduce Gustavsson's algebraic structure. Elements of $\cal B$ are maps from $\cal A $ to itself defined by
\begin{equation}
as_{\alpha,\beta}=[\alpha,\beta,X]\, ,
\end{equation}
where $X\in{\cal A}$. Taking a commutator of two such maps produces another map from $\cal A$ to itself and we have a Lie algebra. Defining the products
\begin{eqnarray}
 \langle\alpha,\beta\rangle &=& as_{ \alpha,\beta} \, ,\\
(as_{\alpha,\beta},\gamma) &=& (as_{\alpha,\beta}(\gamma))\, ,
\end{eqnarray}
it is fairly straightforward to check that the Jacobi-like identities and associativity condition are met.

The theory is always used in the Bagger-Lambert form but is usually referred to as the BLG theory.

\section{D2s from M2s}

Throughout these lectures we have kept an eye on dimensional reduction from M-theory to string theory. We saw how the supermembrane action could be dimensionally reduced to that of a D2 brane, via the abelian dualisation of one of the scalars to a gauge field. Now we are discussing the multiple brane case something more complex must be at work. The theory on the D-branes is non-Abelian and we must get a Yang-Mills kinetic term somehow from the non-propagating gauge field on the membranes which has Chern-Simons action. Mukhi and Papageorgakis showed how to do this with a novel Higgs mechanism\cite{Mukhi:2008ux}. We will show how this is done specifically for the simplest 3-algebra ${\cal A}_4$ that we met above, though the general case proceeds similarly.

We begin by giving a vev to one of the scalars, chosen to be $X^8$, and we can use the $SO(4)$ invariance to point this along a particular 3-algebra direction, 4, which we rename $\phi$ making the split of the 3-algebra indices from $A,B,\ldots\in\{1,2,3,4\}$ to $a,b,\ldots\in\{\{1,2,3\}, \phi\}$. The expectation value is the radius of the M-theory circle, $R$, but in order for the scalar field to have canonical dimension we must divide by a power of the Planck length such that using the M-theory/string theory dictionary we find we have
\begin{equation}
\langle X^{\phi (8)}\rangle = g_{YM}\ .
\end{equation}
Note that giving such a vev to a single scalar still preserves the full supersymmetry.

We concentrate on the reduction of the gauge field. After singling out the $\phi$ direction our 3-algebra gauge field $A_\mu^{AB}$ can be rewritten in terms of two ordinary gauge fields,
\begin{eqnarray}
A_\mu^{\ a}&:=&A_\mu^{\ a\phi},\\
 B_\mu^{\ a}&:=&\frac{1}{2}\epsilon^{a}_{\ bc}A_\mu^{\ bc}.
\end{eqnarray}
We also define a new covariant derivative and field strength in terms of $A_\mu^{\ a}$ only  by
\begin{eqnarray}
D'_\mu X^{a(I)}&=&\partial'_\mu X^{a(I)}-2\epsilon^{a}_{\ bc}A_\mu^{\ b}X^{c(I)},\\
F'_{\nu\lambda}{}^{ a}&=&\partial_\nu A_\lambda^{\ a}-\partial_\lambda A_\nu^{\ a}-2\epsilon^{a}_{\ bc}A_\nu^{\ b}A_\lambda^{\ c}.
\end{eqnarray}
Substitution in the Chern-Simons and scalar kinetic terms yields the following terms involving $B_\mu{}^a$;
\begin{equation}
{\cal L} =  -2 g_{YM}^2 B_\mu^{~a}B^\mu_a -2 g_{YM}B^{~a}_\mu
  D'^\mu X_a^{(8)} + 2\,\epsilon^{\mu\nu\lambda}\,
  B_\mu^{~a} F'_{\nu\lambda a} + \ldots \end{equation}
  The first is a mass term for $B_\mu{}^a$ and the dots denote higher order terms which will be suppressed in the large $g_{YM}$ limit. We see $B_\mu{}^a$ appears without derivatives and can be eliminated using its equation of motion
\begin{equation}
B_\mu^{~a} =
\frac{1}{2g_{YM}^2}
\epsilon_\mu^{\phantom{\mu}\nu\lambda}\,F_{\nu\lambda}^{'a} 
- \frac{1}{2g_{YM}}D'_\mu X^{a(8)}\, .
\end{equation}
We see that the the first term is proportional to the field strength of $A_\mu{}^a$ and that the mass terms for $B_\mu{}^a$ will now give a standard Yang-Mills kinetic term exactly as hoped!
It can be shown that at leading order the Lagrangian is that of $SU(2)$ Yang-Mills, along with the following decoupled terms
\begin{equation}
{\cal L}_{\rm decoupled}= -\frac{1}{2}\partial_\mu X^{\phi(I)}\partial^\mu X^{(I)}_{\phi}
+ \frac{i}{2}{\bar \Psi}^\phi \Gamma^\mu\partial_\mu\Psi_\phi\;.
\end{equation}
There are 8 scalars, but one can be made into a $U(1)$ gauge field via Abelian duality giving an Abelian multiplet. This can be given the interpretation of the centre of mass modes for the D2-brane theory. 

\section{Progress and Problems}

There has been much work on the BLG theory in the last few years, but there are still some issues remaining unresolved. Not least is the lack of knowledge about 3-algebras. We have mentioned only the simplest 3-algebra ${\cal A}_4$, but in fact it is also the most complicated 3-algebra as it has been shown to be the only one!\cite{Papadopoulos:2008sk,Gauntlett:2008uf} Relaxing the constraint of positivity of the norm gets around this\footnote{In fact as it is the positivity of the norm that guarantees unitarity, it is more appropriate to relax the total anti-symmetry of the bracket, which also leads to more 3-algebras (e.g. \cite{Bagger:2008se} ).}, and all Lie algebras and more can then be embedded in so called Lorentzian 3-algebras\cite{Gomis:2008be,Benvenuti:2008bt,Ho:2008ei}  but of course one then has to deal with the negative norm states. Models in which these were shown to vanish\cite{Bandres:2008kj,Gomis:2008uv} turned out to be equivalent to Yang-Mills theory of D2-branes and so are thought not to describe M2-branes at all\cite{Ezhuthachan:2008ch}. This means for all its promise the BLG Lagrangian as described here can only describe the ${\cal A}_4$ theory, and it is still unclear how many membranes this is to be interpreted as. Bagger and Lambert originally postulated that it described three membranes (this was when supplemented by the central element $T^0$ which commuted with everything and was taken describe the centre of mass). That raises the question of why there is no two membrane theory, and Mukhi and Papageorgakis interpret their dimensional reduction, in which the D-brane centre of mass multiplet appears form within ${\cal A}_4$, as indicating there is no need to add this separately, and that the theory better described two membranes. 

Something else we have not mentioned is the quantisation of the structure constants, $f^{abcd}$. In order to be invariant under large gauge transformations in the quantum theory, Chern-Simons terms must always appear with a coefficent $\pi/k$, where $k$ is an integer called the level. In the BLG theory this implies the quantisation of $f^{abcd}$ and there are no continuous free parameters, so it is expected the quantum theory should remain superconformal, there are no coupling constants to run. The question remains how the quantised parameter $k$ is to be interpreted in terms of M-theory branes.

Another outstanding issue was how the action could be parity invariant, as required for multiple membranes, when Chern-Simons terms are not. Attempts to explain this away seemed a little ad hoc. We will see that the ABJM theory, which in general has ${\cal N}=6$ supersymmetry (but with the BLG theory as a special case), gives insight into many of these questions.

\chapter{ABJM}

The ABJM theories are ${\cal N}=6$ superconformal Chern-Simons matter theories with gauge group $SU(N)\times SU(N)$ or $U(N)\times U(N)$. They are interpreted as the worldvolume theory of multiple membranes in certain backgrounds. Before we discuss details of these theories we will see how we can write the BLG theory as a Chern-Simons matter theory, in a presentation due to Van Raaamsdonk (\cite{VanRaamsdonk:2008ft}, see also \cite{Berman:2008be}).

\section{BLG as Bifundamental Gauge Theory}

We saw that the gauge fields of the BLG model for ${\cal A}_4$ take values in $SO(4)$ (here we define ${\cal A}_4$ as $f^{abcd}=f \epsilon^{abcd}$, where the invariance under large gauge transformations will fix $f=2\pi/k$ for integral $k$). Writing $SO(4)$ as $SU(2)\times SU(2)$ a real vector becomes bifundamental as we can see through
\begin{equation}
X^{I} = {1 \over 2} \left( \begin{array}{cc} x^I_4 + i x^I_3 & x^I_2 + i x^I_1 \cr -x^I_2 + i x^I_1 & x^I_4 - i x^I_3 \end{array} \right).
\end{equation}
We decompose the BLG gauge field into its self-dual and anti-self-dual parts obeying $A_{\mu a b}^\pm = \pm {1 \over 2} \epsilon_{abcd} A^\pm_{\mu cd}$ via
\begin{equation}
A_{\mu a b} = - {1 \over 2f} (A^+_{\mu a b} + A^-_{\mu a b})\,. 
\end{equation}
Using the Pauli matrices (with normalisation $\mbox{Tr}(\sigma_i \sigma_j) = 2 \delta_{ij}$) we define gauge fields
\begin{eqnarray}
A_\mu &=& A^+_{\mu 4 i} \sigma_i\, ,\\
 \hat{A}_\mu &=& A^-_{\mu 4 i} \sigma_i \, .
\end{eqnarray}
Rewriting the action these appear with ordinary Chern-Simons terms, but with opposite signs:
\begin{eqnarray}
{\cal L} &=&  \mbox{Tr}( -(D^\mu X^I)^\dagger D_\mu X^I + i \bar{\Psi}^\dagger \Gamma^{\mu} D_\mu \Psi )\cr
&& + \mbox{Tr}(-{2 \over 3} if \bar{\Psi}^\dagger \Gamma_{IJ} (X^I X^{J \dagger} \Psi + X^J \Psi^\dagger X^I + \Psi X^{I \dagger} X^J)  - {8 \over 3}f^2 X^{[I} X^{J \dagger} X^{K]}X^{K \dagger} X^J X^{I \dagger}) \cr
&& + {1 \over 2f} \epsilon^{\mu \nu \lambda} \mbox{Tr}(A_\mu \partial_\nu A_\lambda + {2 \over 3}i A_\mu A_\nu A_\lambda)  - {1 \over 2f} \epsilon^{\mu \nu \lambda} \mbox{Tr}(\hat{A}_\mu \partial_\nu \hat{A}_\lambda + {2 \over 3}i \hat{A}_\mu \hat{A}_\nu \hat{A}_\lambda) \cr
\end{eqnarray}
with covariant derivative $D_\mu X^I = \partial_\mu X^I + i A_\mu X^I -i X^I \hat{A}_\mu$.

The issue of parity becomes clearer, with the action now invariant under parity transformation and the exchange of the gauge fields, $A_\mu \leftrightarrow \hat{A}_\mu$ (we note that parity also takes $X^I\leftrightarrow X^{I \dagger}$).

 \section{BLG in ${\cal N}=2$ Superspace}
 
 The ABJM theories are usually presented in  ${\cal N}=2$ superspace and we will proceed by writing the BLG theory in this way. We first combine the $X^I$ into complex scalars given by
\begin{equation}
Z^A = X^A + iX^{A+4},\qquad A=1,\ldots,4.
\end{equation}
An $SU(4)$ subgroup of the original $SO(8)$ R-symmetry is manifest and acts on the index $A$. These are then combined with the fermions into bi-fundamental chiral superfields $\mathcal{Z}^A$, while the gauge fields are contained in vector superfields $\mathcal{V}$ and $\hat{\mathcal{V}}$. Recall that in component form in Wess-Zumino gauge these are given
\begin{eqnarray}
\mathcal{Z}&=& Z(x_L)
       + \sqrt{2} \theta\zeta(x_L)
       + \theta^2 \, F(x_L) \; , \\ 
       \bar{\mathcal{Z}}&=& Z^\dagger(x_R)
       - \sqrt{2}\bar{ \theta}\zeta^\dagger(x_R)
       + \bar{\theta}^2 \, F^\dagger(x_R) \; , \\ 
\mathcal{V} &=& 2i \, \theta \bar{\theta} \, \sigma(x)
    + 2 \, \theta\gamma^\mu\bar{\theta} \, A_\mu(x)
    + \sqrt{2} i \, \theta^2 \, \bar{\theta}\bar{\chi}(x) 
    - \sqrt{2} i \, \bar{\theta}^2 \, \theta \chi(x)
    + \theta^2 \, \bar{\theta}^2 \, D(x)
\end{eqnarray}
where $x_L^\mu=x^\mu+i\theta\gamma^\mu\bar{\theta},\ x_R^\mu=x^\mu-i\theta\gamma^\mu\bar{\theta}$.

We define conjugations by
\begin{align}
  Z^{\ddagger A} &:=  X^{\dagger A} + i X^{\dagger A+4} \; ,  \\
  \bar{Z}_A      &:=X^A           - i X^{A+4} \; . 
\end{align}
The first of these acts on the $SU(2)$ representations, the second of these on the $SU(4)$ representations, the combination is hermitian conjugation.
The BL Lagrangian can then be written in superspace as the sum of three parts,
\begin{eqnarray}
S_{\mathrm{mat}} &=& - \int d^3x\,d^4\theta\: \mbox{tr} \bar{\mathcal{Z}}_A e^{-\mathcal{V}} \mathcal{Z}^A e^{\hat{\mathcal{V}}} \; ,  \\
S_{\mathrm{CS}} &=& -\frac{i}{4f}  \int d^3x\,d^4\theta \int_0^1 dt\: \mbox{tr} \Bigl[ \mathcal{V} \bar{D}^\alpha \Bigl( e^{t \mathcal{V}} D_\alpha e^{-t \mathcal{V}} \Bigr) - \hat{\mathcal{V}} \bar{D}^\alpha \Bigl( e^{t \hat{\mathcal{V}}} D_\alpha e^{-t \hat{\mathcal{V}} } \Bigr)\Bigr] \; ,\\
S_{\mathrm{pot}} &=& 4f \int d^3x\,d^2\theta\: \mathrm{W}(\mathcal{Z}) + 4f \int d^3x\,d^2\bar{\theta}\: \bar{\mathrm{W}}(\bar{\mathcal{Z}}) \, ,
\end{eqnarray}
with the superpotential given by
\begin{eqnarray}
 \mathrm{W}      &=& \frac{1}{4!} \epsilon_{ABCD} \mbox{tr}\mathcal{Z}^A \mathcal{Z}^{\ddagger B} \mathcal{Z}^C \mathcal{Z}^{\ddagger D},\\
   \bar{\mathrm{W}} &=& \frac{1}{4!} \epsilon^{ABCD}\mbox{tr}\bar{\mathcal{Z}}_A \bar{\mathcal{Z}}_B^\ddagger \bar{\mathcal{Z}}_C \bar{\mathcal{Z}}_D^\ddagger
  \; ,\label{nbc:blgW}
\end{eqnarray}
which has manifest $U(1)_R \times SU(4)$ global symmetry. This was shown by Benna, Klebanov, Klose and Smedback\cite{Benna:2008zy} and can be checked by writing out in components and integrating out all auxiliary fields.

\section{Generalising Away from $SU(2)$}

The ABJM theory can be thought of as a generalisation of the this theory away from $SU(2) \times SU(2)$. We give up the manifest $SU(4)$ invariance  by relabelling $Z^3\rightarrow W_1$ and $Z^4\rightarrow W_2$, with corresponding chiral superfield expressions; our bifundamental matter superfields are now $\mathcal{Z}^1,\mathcal{Z}^2, \mathcal{W}_1, \mathcal{W}_2$. We can write the BLG superpotential (\ref{nbc:blgW}) as
\begin{eqnarray}
 \mathrm{W}      &=& \frac{1}{4} \epsilon_{AC}\epsilon^{BD} \mbox{tr}\mathcal{Z}^A \mathcal{W}_{B} \mathcal{Z}^C \mathcal{W}_{ D},\\
   \bar{\mathrm{W}} &=& \frac{1}{4} \epsilon^{AC}\epsilon_{BD}\mbox{tr}\bar{\mathcal{Z}}_A \bar{\mathcal{W}}^B \bar{\mathcal{Z}}_C \bar{\mathcal{W}}^D
  \; .
\end{eqnarray}
Manifest global symmetries are now $SU(2)\times SU(2)$ (part of the R-symmetry, not to be confused with the gauge symmetry) as well the $U(1)_b$ baryonic symmetry
\begin{equation}
\mathcal{Z}^A\rightarrow e^{i \alpha}\mathcal{Z}^A,\qquad\quad\mathcal{W}_B\rightarrow e^{-i\alpha}\mathcal{W}_B\, .
\end{equation}
The conjugation $\ddagger$ was particular to the $SU(2)\times SU(2)$  gauge group, and now the superpotential has been rewritten without it, we can generalise to gauge groups $U(N)\times U(N)$ with $\mathcal{Z}$ in the $(\mathbf{N},\bar{\mathbf{N}})$ and $\mathcal{W}$ in the $(\bar{\mathbf{N}},\mathbf{N})$. This gives the ABJM model\cite{Aharony:2008ug}. Although only $U(1)\times SU(2)\times SU(2)$ global symmetry is manifest, the ABJM model actually has $SU(4)\sim SO(6)$ R-symmetry and ${\cal N} =6$ supersymmetry. We write $\mathcal{W}$ in component fields as.
  \begin{equation}
   \mathcal{W} = W(x_L)
       + \sqrt{2} \theta\omega(x_L)
       + \theta^2 \, G(x_L)\, .
\end{equation}
We then write the scalars and fermions in the $SU(4)$ combinations 
\begin{equation}
  Y^A = \{ Z^A, W^{\dagger A} \}
  , \qquad Y_A^\dagger = \{ Z_A^\dagger, W_A \} \; ,
  \end{equation}
and
\begin{eqnarray}
  \psi_A &=& \{ \epsilon_{AB} \zeta^B \, e^{-i\pi/4}, -\epsilon_{AB} \omega^{\dagger B} \, e^{i\pi/4} \}\, , \\
  \psi^{A\dagger}& =& \{ -\epsilon^{AB} \zeta_B^\dagger \, e^{i\pi/4}, \epsilon^{AB} \omega_B \, e^{-i\pi/4} \} \; .
\end{eqnarray}
 The purely bosonic part of the potential is
\begin{eqnarray}
  V^{\mathrm{bos}} = - \frac{f^2}{3} \mbox{tr} \Bigl[
          Y^A Y_A^\dagger Y^B Y_B^\dagger Y^C Y_C^\dagger 
      +   Y_A^\dagger Y^A Y_B^\dagger Y^B Y_C^\dagger Y^C
\\ \hspace{11mm}
      + 4 Y^A Y_B^\dagger Y^C Y_A^\dagger Y^B Y_C^\dagger 
      - 6 Y^A Y_B^\dagger Y^B Y_A^\dagger Y^C Y_C^\dagger 
   \Bigr] \; .
\end{eqnarray}
while the Fermionic pieces also have a manifestly  $SU(4)$ invariant form
\begin{eqnarray}
   V^{\mathrm{ferm}} &=& \frac{i L}{4} \mbox{tr} \Bigl[
        Y_A^\dagger Y^A \psi^{B\dagger} \psi_B
      - Y^A Y_A^\dagger \psi_B \psi^{B\dagger}
      + 2 Y^A Y_B^\dagger \psi_A \psi^{B\dagger}\\
       &&- 2 Y_A^\dagger Y^B \psi^{A\dagger} \psi_B
    - \epsilon^{ABCD} Y_A^\dagger \psi_B Y_C^\dagger \psi_D
      + \epsilon_{ABCD} Y^A \psi^{B\dagger} Y^C \psi^{D\dagger}
      \Bigr] \; .
\end{eqnarray}

\section{What Does ABJM Describe?}

The BLG theory was supposed to describe membranes in in flat space, with the required ${\cal N}=8$ supersymmetry, but just how many membranes it described was unclear, how exactly were $N$ and $k$ to be interpreted? And now we have a theory with ${\cal N}=6$, what does it describe?

Let's look at the moduli space of the theory. In the $U(1)\times U(1)$ case the superpotential and Bose-Fermi couplings vanish and we have a free theory of the four superfields $Y^I$. At first glance the moduli space is just ${\mathbb C}^4$, but we have to be more careful with the gauge fields. We have standard gauge transformations $A\rightarrow A-d\Lambda$, $\hat{A}\rightarrow \hat{A}-d\hat{\Lambda}$ and we can gauge fix $A$ to zero, but then there still remain large gauge transforms with the $\Lambda$'s everywhere constant. In the presence of a boundary, the Abelian Chern-Simons action, which has the form $A\wedge dA$, is not invariant under such gauge transforms. We have that
\begin{equation}
\delta S_{CS}=\frac{k}{2\pi}\int_{\partial {\cal M}}(\Lambda\wedge F -\hat{\Lambda}\wedge\hat{F}).
\end{equation}
Over any 2-manifold the integral of the the field strengths are quantised, $\int F \in 2\pi{\mathbb Z}$. In order that the path-integral be invariant the Chern-Simons action must transform by a multiple of $2\pi i$, so we require that $\Lambda=2\pi n/k$ with $n\in {\mathbb Z}$ and similarly for $\hat{\Lambda}$. As the scalars transform as $Y_I\rightarrow 2^{i(\Lambda-\hat{\Lambda})}Y_I$ under gauge transforms we see that the moduli space is not ${\mathbb C}^4$ but ${\mathbb C}^4/{\mathbb Z}_k$, where the ${\mathbb Z}_k$ symmetry acts as  $Y_I\rightarrow 2^{2\pi i/k}Y_I$. The $SU(4)$ symmetry is manifest.

For the $U(N)\times U(N)$ case the scalar potential vanishes for diagonal $Y_I$, and in fact this is the full moduli space of the theory. The gauge symmetry is thus broken to $U(1)^N\times U(1)^N  \times S_N$, with $S_N$ permuting the diagonal elements.  Things proceed as before and the moduli space for general $N$ is $({\mathbb C}^4/{\mathbb Z}_k)^N/S_N$. The $SU(N)\times SU(N)$ case can also be analysed, and the result is a slightly more complicated orbifold of ${\mathbb C}^4$.

Now this is the moduli space of $N$ M2-branes probing a ${\mathbb C}^4/{\mathbb Z}_k$ singularity, and this is how ABJM interpreted their theory. Note that this membrane theory has the $SU(4)\times U(1)$ isometry surviving from the original $SO(8)$ and so generically has ${\cal N}=6$ supersymmetry.

The spinors which were in the $\mathbf{8}_c$ of $SO(8)$ and decompose into $SU(4)\times$ $U(1)$ reps as $\mathbf{6}_0+\mathbf{1}_2+\mathbf{1}_{-2}$. The last two supercharges are projected out by the orbifold if $k>2$, but for $k=1,2$ the theory should have ${\cal N}=8$ supersymmetry. Another way to see this is that the spinors transform as 
\begin{equation}
\Psi \rightarrow e^{2\pi i (s_1+s_2+s_3+s_4)/k}\Psi
\end{equation}
under the action of the ${\mathbb Z}_k$, where $s_i = \pm 1/2$ are the spinor weights. The chirality projection also requires the sum of the $s_i$ to be even. There are six choices leaving the spinor invariant with the $s_i$ summing to zero, when $k=1$ or $k=2$ they can also sum to $2$ giving two additional spinors and we have ${\cal N}=8$ supersymmetry\footnote{A clear exposition of this can be found in Section 3.2.2 of \cite{Berman:2009xd}.}

As well as these moduli space arguments, ABJM were able to provide a brane construction of their field theory which in the IR limit of the M-theory lifted to membranes probing the singularity at the intersection of two Kaluza-Klein monopoles, which is exactly the same background as above. 

To conclude ABJM found a superconformal field theory with ${\cal N}=6$ and bifundamental matter in $U(N)\times U(N)$ (or $SU(N)\times SU(N)$) where the two Chern-Simons gauge fields have equal and opposite levels $k$ and $-k$. The $U(N)\times U(N)$ theory is believed to describe $N$ M2-branes in an ${\mathbb C}^4/{\mathbb Z}_k$ background. For $N=2$ it is equivalent to the BLG theory and has ${\cal N}=8$ supersymmetry\footnote{The are many subtleties to this relation, see the recent discussion in \cite{Lambert:2010ji}}, and for $k=1,2$ the supersymmetry is also enhanced to ${\cal N}=8$ and the membranes are in flat space and ${\mathbb C}^4/{\mathbb Z}_2$ respectively.

\section{Supersymmetry Enhancement and Monopole Operators}

The key to the enhancement to ${\cal N}=8$ supersymmetry for $k=1,2$ is monopole operators\cite{Borokhov:2002ib}. This was already noted in \cite{Aharony:2008ug} and is explained in\cite{Klebanov:2009sg}.

For $SO(8)$ R-symmetry we must have 28 conserved currents. For ABJM we have manifest $U(1)_b\times SU(4)_R$ global symmetry. We get 15 conserved traceless $SU(4)$ currents given by
\begin{equation}
j^{A}_{\mu B} = i\mbox{Tr} \left[Y^A \mathcal{D}_{\mu} Y^\dagger_B - (\mathcal{D}_{\mu} Y^A) Y^\dagger_B + i \psi^{\dagger A} \gamma_{\mu} \psi_B\right].
\end{equation}
The $U(1)_b$ current is related by the $A^-$ equation of motion (where for $U(N)\times U(N)$ theory, $A^\pm \sim \mbox{Tr}A \pm\mbox{Tr}\hat A$) to the current
$
j_\mu \sim \epsilon_{\mu \nu \lambda} F^{+\nu\lambda}\ .
$
So to carry baryonic charge a field configuration must have a flux of $F^+$ through the $S^2$ at infinity - such a flux is created by monopole operators. Monopole operators create a quantised flux in a $U(1)$ factor of the gauge group through a sphere surrounding the insertion point. For $U(N)\times U(N)$ these are labelled by $2N$ integers $q_i,\hat{q}_i$ which can be arranged to be decreasing. They transform in representations corresponding to Young tableaux, the row lengths given by $q_i$ times the Chern Simons level $k$. Thus for $k=1$ or $k=2$ there are monopole operators $({\cal M}^{-2})_{ab}^{\hat{a}\hat{b}}$ which have $kq_1=k\hat{q}_1=2$. These can be combined with the 6 antisymmetric currents
\begin{equation}
j^{AB}_{\mu} = i\left [ Y^A\mathcal{D}_{\mu} Y^B -\mathcal{D}_{\mu} Y^A Y^B + i \psi^{\dagger A} \gamma_{\mu} \psi^{\dagger B}\right ].
\end{equation}
to give currents which are conserved and gauge invariant. The are a further 6 complex conjugates (combining $j_{AB\mu}$ with ${\cal M}^2$) giving an enhancement to total of 28 currents for these values of $k$ only.

\section{Gravitational Dual}

The field theory on multiple membranes has a gravitational dual, just like the more familiar $AdS_5\times S^5$ correspondence. Placing M2-branes at the singularity of an ${\mathbb C}^4/{\mathbb Z}_k$  orbifold the extremal geometry is
\begin{eqnarray}
d s^2_{11} &=& h(\rho)^{-2/3}\left(-dt^2 + dx^2_1 + dx^2_2\right) + h(\rho)^{1/3}\left(dr^2 + \rho^2d\Omega^2_7\right),\nonumber \\
h(r) &=& 1 + \frac{R^6}{\rho^6}\ ,\qquad R^6 = 32 \pi^2 N' l^6_p,\nonumber \\
F_4 &=& d^3x \wedge dh(\rho)^{-1},
\end{eqnarray}
 where $d\Omega^7=S^7/{\mathbb Z}_k$. The near horizon geometry is $AdS_4\times S^7/{\mathbb Z}_k$. The $k=1$ case has been extensively used a basis for compactifications. The 't Hooft coupling is given by $\lambda=N/k$, the field theory is weakly coupled for $k\ll N$. Using the Hopf fibration of $S^7$ as an $S^1$ fibration over ${\mathbb CP}^3$ and the ${\mathbb Z}_k$ quotient acts on the $S^1$ to make it smaller. After the modding out  of the circle by $k$ we see that with our precious conventions $N'=Nk$, and the radius of the circle is $R/k\sim (N/k^5)^{1/6}$. We take this circle as the M-theory circle, and in order for it to remain large so that an M-theory description is valid  we require $k^5\ll N$. For larger $k$ we can use type IIA strings on $AdS_4\times{\mathbb CP}_3$.
As usual for an $AdS$/CFT correspondence we can match chiral operators between the field theory side and the gravitational dual. Once again monopole operators are required to complete the matching. 
The arguments of previous lectures about black brane thermodynamics persist and the gravity dual indicates there should be $N^{3/2}$ degrees of freedom. There is still no understanding of  this on the field theory side.
 
 \section{Outlook}
 
 Immediately following the BLG and ABJM papers there was a huge cascade of literature on the subject (which we have not attempted to review here), and although this may have slowed now, there are still many interesting papers being published. While ABJM has been generalised in many ways and since all mention of three-algebras is gone from that formulation this has tended to take the focus off the appearances of some new kind of algebra, perhaps involving non-associativity. However, Bagger and Lambert have shown that ABJM can be written in terms of a modified 3-bracket and algebra that does not have all the antisymmerty of the original\cite{Benna:2008zy}. There has also been work on trying to understand the $M5$ brane using three-algebras as well\cite{Lambert:2010wm}. However, there seems there is still much to be done to fully understand the mysterious M-theory membrane, and indeed M-theory itself. 

\section*{Acknowledgements}
I would like to thank Dan Thompson for reviewing a draft of these notes, and the participants of the Modave School for helpful questions. This work is supported in part by the Belgian Federal Science Policy Office through the Interuniversity Attraction Pole IAP VI/11 and by FWO-Vlaanderen through project G011410N.


\cleardoublepage
\pagestyle{plain}
\def\href#1#2{#2}
\bibliographystyle{BibliographyStyle}
\addcontentsline{toc}{chapter}{\sffamily\bfseries Bibliography}

\bibliography{BibliographyModave2010}

\end{document}